\documentclass[10pt,conference]{IEEEtran}
\pdfoutput=1
\usepackage[pdftex]{graphicx}
\usepackage{multirow}
\usepackage{balance}
\usepackage{color}
\usepackage{balance}

\begin{document}
\title{Evaluation of Handover Exchange Schemes Between Two Cognitive Radio Base Stations with and without Buffers}
\author{\IEEEauthorblockN{
		Ebenezer Esenogho\IEEEauthorrefmark{1},  
		Elie N. Mambou\IEEEauthorrefmark{2},  
	} 
	\IEEEauthorblockA{\IEEEauthorrefmark{1}
		Centre for Telecommunication, Dept. of Electrical and Electronic Engineering Science, University of Johannesburg, \\P. O. Box 524, Auckland Park, 2006, South Africa, ebenezere@uj.ac.za}
	\IEEEauthorblockA{\IEEEauthorrefmark{2}% 2nd affiliations
		Centre for Telecommunication, Dept. of Electrical and Electronic Engineering Science, University of Johannesburg, \\P. O. Box 524, Auckland Park, 2006, South Africa, emambou@uj.ac.za}
}
\maketitle
\begin{abstract}
This article investigates and evaluate a handover exchange scheme between two secondary users (SUs) moving in different directions across the handover region of neighbouring cell in a cognitive radio network. More specifically, this investigation compares the performance of SUs in a cellular cognitive radio network with and without channel exchange systems. The investigation shows reduced handover failure, blocking, forced and access probabilities respectively, for handover exchange scheme with buffer as compared to the other scenario. 
\end{abstract}
\section{Introduction}\label{sec1}
Spectrum handover is an area of interest in cognitive radio networks presently. Handover is the passage of request/call from one user to the other. A spectrum handover occurs when the primary users (PUs) and SUs collide on the same spectrum hole \cite{wang2011}. The collision could still be among SUs especially when the PUs has vacated the spectrum. Therefore, efficient radio resource in cognitive cellular network is essential to determine the Quality of Service (QoS). Key metrics for evaluating the QoS are but not limited to; the handover call blocking and dropping/handover failure probability respectively \cite{nguyen2012}. Cognitive cellular networks comprise of several cells in which its sizes depend on the physical area. High-density areas require smaller sizes than larger sizes thus, the reason is to accommodate the large number of SUs. Large cell size often causes more blocking for two class of SUs (real and non-real-time traffic) due to the frequent hand-off from one cell to another \cite{onyishi2013}. A well-planned handover procedure is crucial in sustaining continuity of ongoing calls. However, maintaining minimum likelihood of dropping/blocking new calls, processing and traffic exchanging on the network is a challenge. Hence, it is worthwhile to investigate and compare strategies that eases the undesired dropping of calls, reduces the signalling traffic on the network and invariably improve the QoS of the network against schemes without no-exchange (no handover).
\section{Handover types and decision}\label{sec2}
There are two types of handover, which could be precisely categorized as soft and hard handover respectively. A hard handover is an ``open before close'' type of configuration controlled by the mobile switching centre (MSC). The cognitive radio base station (CRBS) transfers the SU’s request to alternative cell and hands off the request. In this type of protocol, the connection preceding the CRBS is ended earlier or as the user is reassigned to the new cell’s CRBS. The SU is linked to not more than one CRBS at any given time \cite{salih2006, cai2010} though, in this investigation, two CRBS were considered. In these schemes, individual users are allocated to one or several channels to avoid channel meddling.  When a SU changes direction between two CRBS, it becomes difficult to mutually interconnect with CRBS because it is making use of separate band. 

In soft handover protocol, several connections can be established with neighbouring and adjacent cells. The nested handover is faced with several challenges. Particularly to develop handover protocols for all-inclusive and to solve these challenges, fast moving  SUs are allotted to the macro cell while pedestrian to the micro cell.  However, macro cell can still serve low speed SUs depending on the present load each SU is carrying at that pointing time. The received signal to noise ratio seen by two neighbouring CRBS determines the handover decision \cite{ni2011}. Furthermore, handover, decision process could be centrally controlled or otherwise thus could be classified as network controlled handover, mobile assisted handover and the mobile controlled handover \cite{harsha2014, alejandro2015}. 

The rest of the paper is organised as follows: Section II briefly discussed handover types and decisions. Section III summarizes related work, Section IV presents the system model of the for the handover exchange, Section V deals with performance measures while numerical results and discussion is in Section VI. The paper is concluded in Section VII.
\section{Related work}\label{sec3}
Several handovers prioritizing procedures are deployed to reduce the forced termination of ongoing calls. However, this result to rise in new call blocking probability and the total accepted traffic \cite{chhotray2014}. \cite{miyim2014} developed a swift and smart nested network layer controller as a new handover strategy to select the optimal network among other networks, while QoS, delays and improved spectral efficiency can be attained. In \cite{malathy2016}, a buffer regime using guard channels is employed such that, both new calls and handover calls are queued. In this scenario, a number of guard channels are reserved for handover calls and when the new calls are congested, a channel from the guard channels is used if it is available. The key contributions of this work are:  
\begin{itemize} 
	\item To compare the performance of a channel handover exchange scheme with queue against no-channel handover exchange scheme through a simulation framework in a cognitive radio network scenario when the PUs are absent.
	\item To investigates the impact of a queuing regime on access/admission probability. In other words, if the queue size decreases, SU access/admission will be affected. 
\end{itemize}

\section{Network and system model for handover exchange}\label{sec4}
\begin{figure}[h]
	\centering
	\includegraphics[width=1\linewidth]{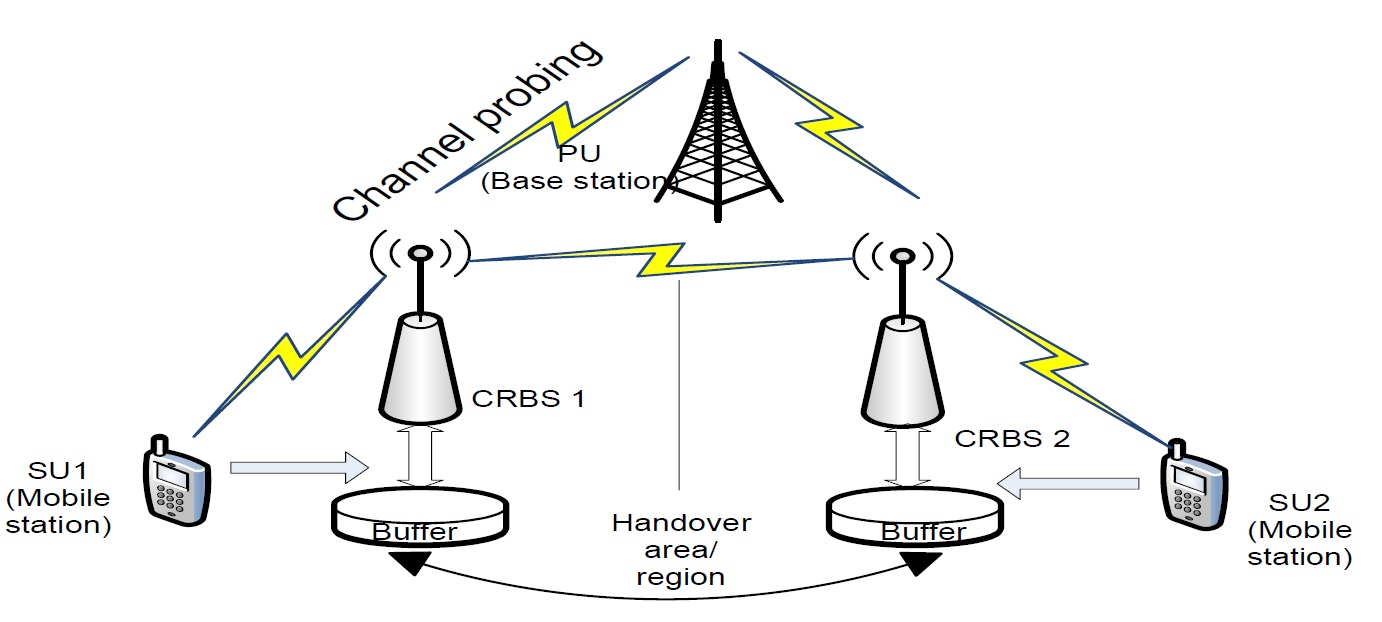}		
	(a)	
	\vspace{8pt}
	\includegraphics[width=1\linewidth]{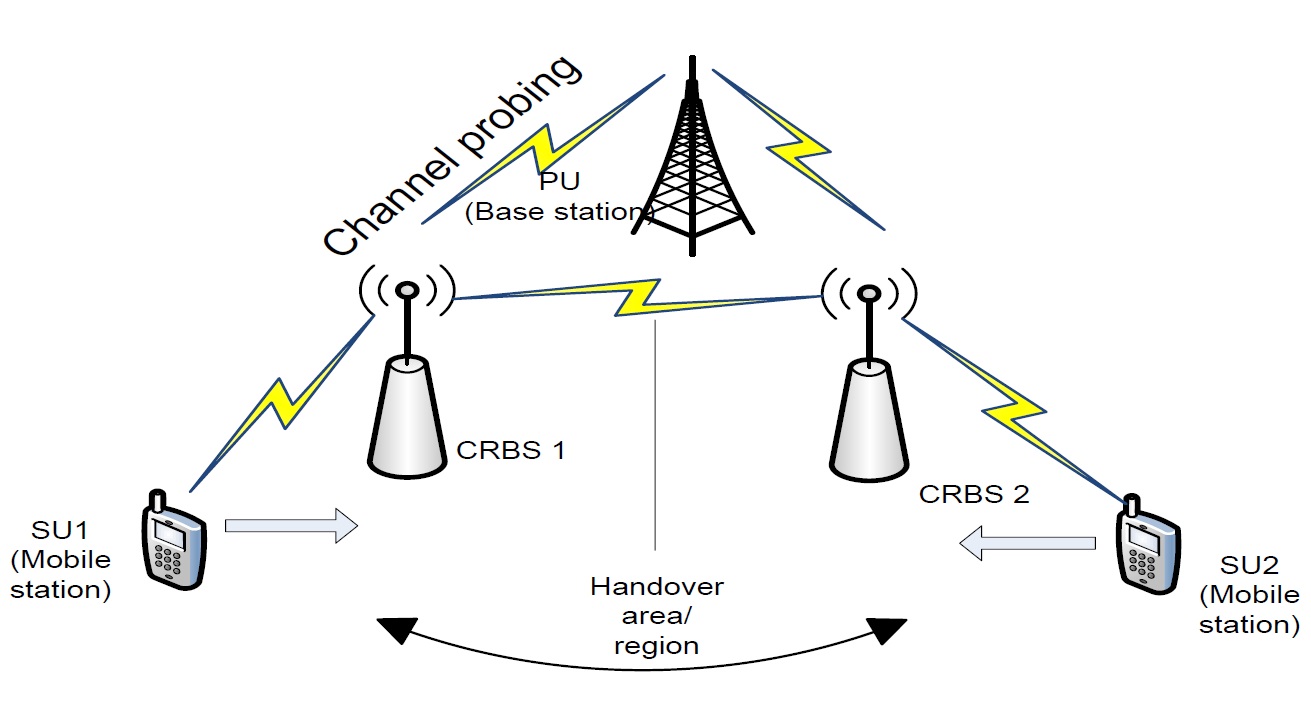}
	(b)		
	\caption{Network model for the handover between CRBS, (a) with buffer and (b) without buffers.}
	\label{fig:fg1}
\end{figure}

The following assumptions where made:
\begin{enumerate}
	\item The channel sensing is perfect and accurate.
	\item The interaction/transaction process is between two CRBS within a define coverage area (a cell).
	\item As at the time of this interaction/transaction, the PUs is absent from the spectrum.
	\item A slow arriving PUs.
	\item SU is used interchangeably used with mobile station (MS) as shown in Fig. \ref{fig:fg1}.
	\item The holding time is exponentially distributed.
\end{enumerate}
 
\begin{figure}[h]
	\centering
	\includegraphics[width=1\linewidth]{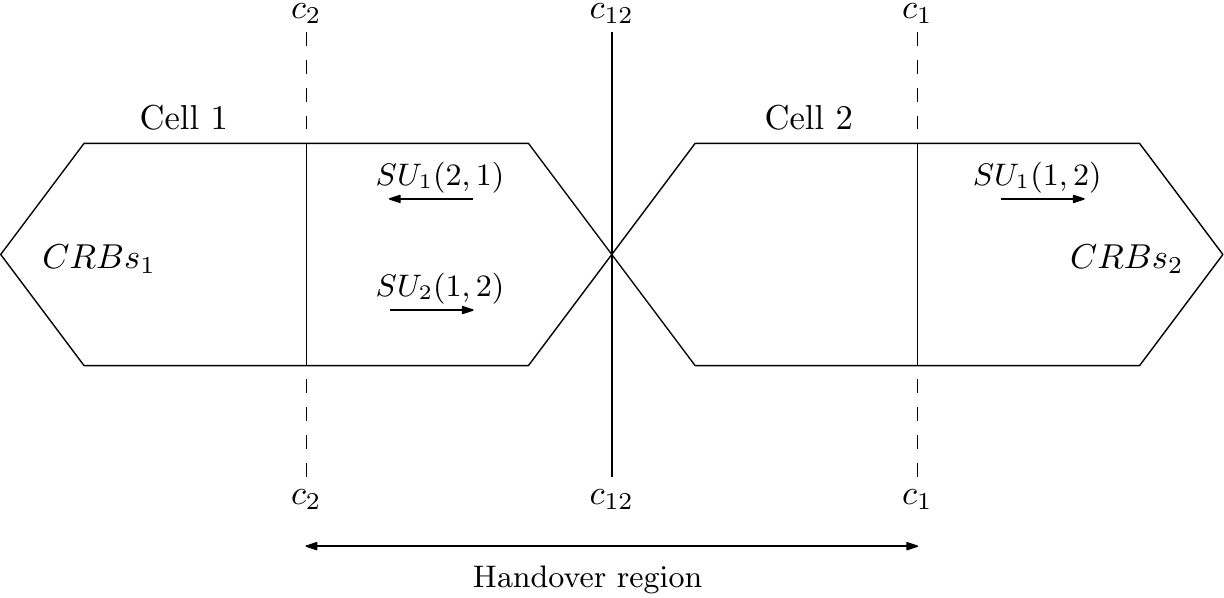}
     \caption{Diagram illustrating of handover region between CRBS cell.}
	\label{fig:fg2}
\end{figure}

The system model consists of a handover area between neighbouring cells shown in Fig. \ref{fig:fg1} Both base stations are separated through the borderline $c_{1,2}$, with lines $c_1$ and $c_2$, represent the assignment region. Outside the right line $c_1$, the signal strength received by a mobile SU from CRBS is inadequate to assure good connections. The symbol  $SU_i(j,k)$ indicates that, the $i^{th}$ SU utilizing a frequency owned by $CRBS_{j}$ and approaching another base station $CRBS_{k}$. Assuming that in an event corresponding to Fig. \ref{fig:fg2}, where both $CRBS_{1}$ and $CRBS_{2}$ channels are occupied. If the condition remains same after a duration, and the $SU_1 (2,1)$ trespasses  $b_2$, a handover failure will occur and the channel hands off in $CRBS_{2}$. 
The $CRB_{s2}$  will formerly allocate this frequency to the $SU_2(1, 2)$, if it is within the handover region. The mobile SU are serviced in the same vain in a traditional resource sharing policy. However, in the handover exchange scheme,  the SU mobiles are permitted to interchange their resources when travelling in reverse directions within handover space. Therefore, with Fig. \ref{fig:fg2}, the channels held by SUs (mobiles), $SU_2(1,2)$ and $SU_1(2,1)$ are exchanged. 
This results in handover success for both SUs (mobiles) unlike the conventional scheme. In channel handover exchange scheme, there is a mutual interaction between neighbouring cells such that, $CRBS_i$  buffer handover calls from a SU travelling from $CRBS_i$ to $CRBS_j$  in different duffer $q_{ij}$. Also, the $CRBS_j$ maintains a queue  $q_{ji}$ for handover request from SUs travelling from region $i$ to $j$.

\begin{figure}[h]
	\centering
	\includegraphics[width=1\linewidth]{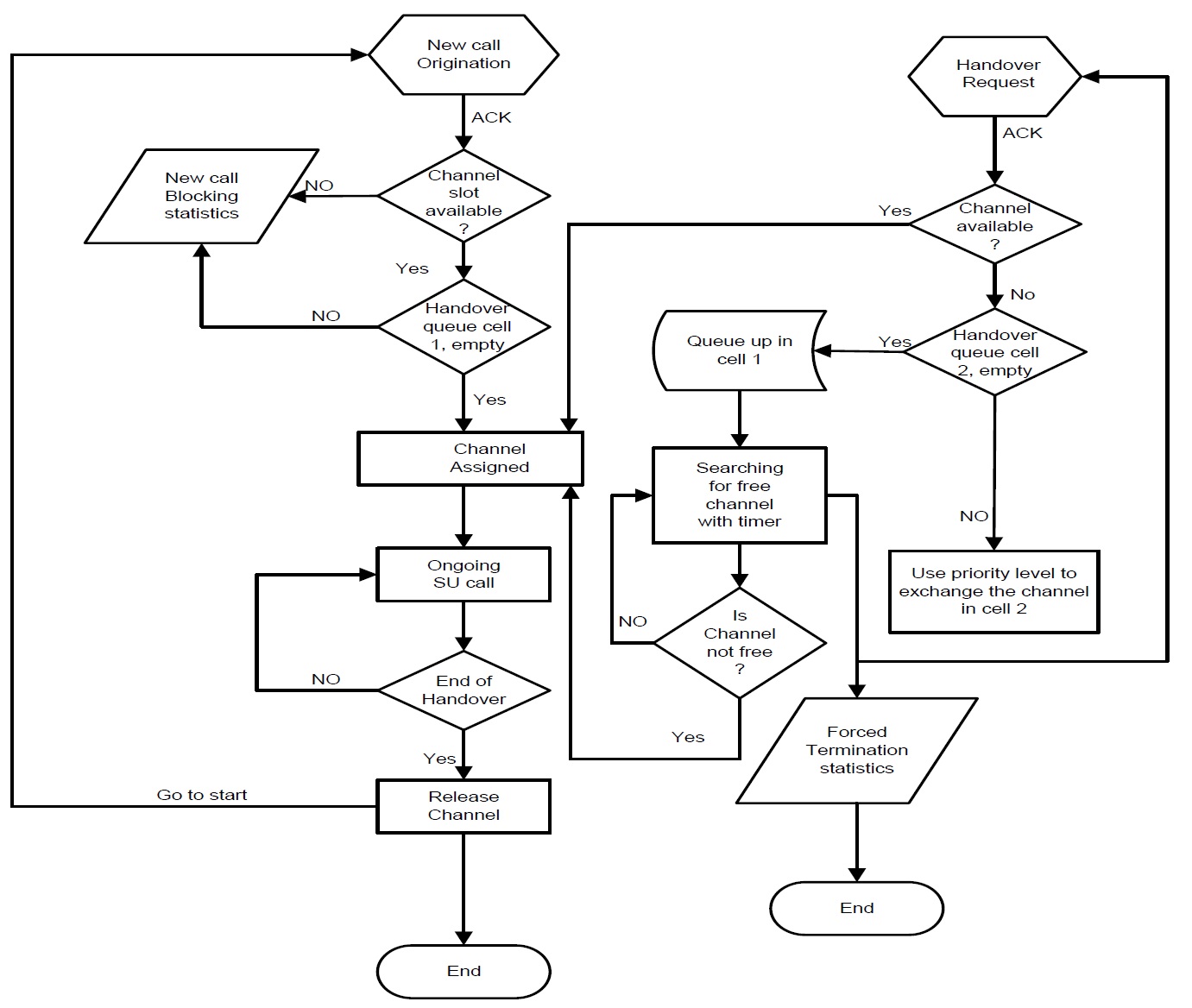}
	\caption{Flow chart of the handover protocol.}
	\label{fig:fg3}
\end{figure}

Both $CRBS_i$ and 〖$CRBS_j$ together blocks request/calls if free channels are not available. However, idle frequency is allocated to new request if the buffer is unfilled. In this scheme, request/calls in the buffer are serviced with priority hence, both buffers will always be filled. The procedure for hand over from the mobile user in cell $i$ to the CRBS of cell $j$ is captured Fig. \ref{fig:fg3} in the following ways:
\begin{enumerate}
	\item When an idle frequency is found in cell $j$, it is allocated to the SU seamlessly; this outcome is a successful handover.
	\item If an idle channel slot is occupied in cell $j$ and buffer $q_{ij}$  is unoccupied, then the handover request is pushed to the buffer in $q_{ij}$ for later service.
	\item If an idle channel slot is occupied in cell $j$ and $q_{ij}$ is full, then the SU is forced to exchange its channel with the channel held by the SU whose, handover request has the top priority in buffer $q_{ij}$.
\end{enumerate}

As soon as the call enqueued, a handover call translates to successful delivery of the calls, when a slot (channel) is handed off in the equivalent coverage area or when a slot is swapped by a SU of that cell as shown in Fig \ref{fig:fg3}. Nevertheless, if a channel is not vacant to a handover as shown in Fig. \ref{fig:fg2}, during the crosses of the handover region, its effects is a handover failure. However, queued handover calls are occasionally ranked as a result of the signal to noise ratio (SNR) from the SUs (mobiles). When a free channel-slot is vacant, it is allocated to the highest urgent SU (mobile). 

Likewise, if a priority channel interchange commenced in the buffer. Delayed handover calls, in the queue which link to handover failures are intermittently removed from the buffer stack. However, here are the three procedures in handover exchange. Firstly, SUs whose call attempts are rejected depart and never come back. Secondly, SU request dropped attempt over again within a predefine time, and lastly, the network permits secondary users to be added to a queue in a buffer until slot are allocated.
\section{Performance measures}\label{sec5}
The terminologies for this simulation are as follows:
\begin{enumerate}
	\item The dwelling time $t_{dl}$ is the duration within which, a call continues without any handover exchange. 
	\item Holding time  $t_h$ is the entire duration a call can last. The rest are found in Table II.
\end{enumerate}
Let $P$ be the likelihood that the $t_{dl}$ of a request is allotted. A channel terminates before its call $t_h$. Then it can be expressed as 
\begin{equation}
P=\frac{\mu_d}{\mu_d+\delta_h}
\end{equation}
Note that, a request is exchanged only when $t_h>t_{dl}$. The likelihood of a new call been successful is $(1-P_{nc})$. So, $\lambda_{nc}(1-P_{nc})$ is the ratio to which incoming request are allocated. Assuming an incoming call is created in cell $i$, the likelihood that is successful is $(1-P_{nc})$. It is formerly exchanged to cell 2, if $t_h>t_{dl}$ with a likelihood of $(1-P_{nc})P$.

\section{Flow chart/algorithm}
This section presents the simplified pseudocode for this system model followed by the algorithm 

\begin{table}\label{tb1}
	\caption{System model algorithm}
	\begin{tabular}{cl}
		\hline
		&\textbf{Pseudo code for new call origination} \\
		\hline
		1.&\textit{New call initiates and acknowledge}  \\ 
		2.&\textit{Is channel slot available}  \\
		3.&\textit{\mbox{ }\mbox{ }\mbox{ }If no} \\
		4.&\textit{New call blocking statistics initiated}  \\
		5.&\textit{\mbox{ }\mbox{ }\mbox{ }If yes (Else)} \\ 
		6.&\textit{Is handover queue in cell 1 empty}  \\
		7.&\textit{\mbox{ }\mbox{ }\mbox{ }\textbf{Go step 4 (If no)}} \\
		8.&\textit{\textbf{Go step 5}} \\
		9.&\textit{\mbox{ }\mbox{ }\mbox{ }\textbf{Go step 6 (If yes)}} \\
		10.&\textit{Allocate/Assign channel}  \\
		11.&\textit{Ongoing SU calls/request}  \\
		12.&\textit{Is handover ended}  \\
		13.&\textit{\mbox{ }\mbox{ }\mbox{ }\textbf{Go step 4}} \\
		14.&\textit{\mbox{ }\mbox{ }\mbox{ }\textbf{Go step 6}} \\
		15.&\textit{Release the channel}  \\
		16.&\textit{\mbox{ }\mbox{ }\mbox{ }\textbf{Go step 1}} \\
		17.&\textit{Else end}  \\
		\hline 
		&\textbf{Handover request procedure}\\
		\hline
		18.&\textit{Handover request and acknowledge}  \\
		19.&\textit{\textbf{Go step 3}} \\
		20.&\textit{\mbox{ }\mbox{ }\mbox{ }\textbf{Go step 6}} \\
		21.&\textit{\textbf{Go step 11}} \\
		22.&\textit{\mbox{ }\mbox{ }\mbox{ }\textbf{Go step 4}} \\
		23.&\textit{Is handover queue in cell 2 empty}  \\
		24.&\textit{\mbox{ }\mbox{ }\mbox{ }\textbf{Go step 4}} \\
		25.&\textit{Use priority level to exchange the channel in cell 2}  \\
		26.&\textit{\mbox{ }\mbox{ }\mbox{ }\textbf{Go step 6}} \\
		27.&\textit{Queue up in cell 1}  \\
		28.&\textit{Start searching/probing for free channel with timer set}\\
		29.&\textit{\textbf{Go step 3}} \\
		30.&\textit{\mbox{ }\mbox{ }\mbox{ }\textbf{Go step 4}} \\
		31.&\textit{\textbf{Go step 29}} \\
		32.&\textit{\mbox{ }\mbox{ }\mbox{ }\textbf{Go step 6}} \\
		33.&\textit{\textbf{Go step 11}} \\
		34.&\textit{If no free channel slot after probing}\\
		35.&\textit{Drop/forcibly terminate or \textbf{Go to step 18}}\\
		36.&\textbf{Else, end}\\
		\hline
	\end{tabular} 
\end{table}

\begin{table}\label{tb2}
	\caption{Performance parameters}
	\centering
	\begin{tabular}{|c|l|}
		\hline
		$\lambda_{nc}:$&New (incoming) call arrival rate.\\
		\hline
		$\lambda_{hoc}:$&Handover call arrival rate.\\
		\hline
		$\mu_{d}^{-1}:$&Average cell dwelling time.\\
		\hline
		$\delta_{h}^{-1}:$&Average cell-holding time.\\
		\hline
		$P_{nc}:$&Incoming call blocking probability.\\
		\hline
		$P_{hf}:$&Handover failure probability.\\
		\hline
			$\mu_{d}:$&Average call duration.\\
		\hline
	\end{tabular}
\end{table}
Therefore, probability that access is granted or successful to a handover is expressed as
\begin{equation}
(1-P_{hf})
\end{equation}
Thus, the likelihood of an incoming request from cell $i$ to successfully exchange with cell $j$ is expressed as 
\begin{equation}
(1-P_{nc})(1-P_{hf})P
\end{equation}
On bases that a call can be exchanged numerous time before dropping, then $\lambda_{hoc}$ can be expressed as
\begin{equation}
\lambda_{hoc}=\frac{\lambda_{nc}(1-P_{nc})P}{(1-P_{nc})P}
\end{equation}
The carried traffic $\alpha_{ct}$, is the simultaneous call supported by the network or cell. It is mostly estimated as the average of an interval which could be an hour or less. In this paper, it denotes a portion of the accessible traffic $\omega_{ot}$. It is expressed as 
\begin{equation}
\alpha_{ot}=\omega_{ot}(1-P_{nc})
\end{equation}                                                               
An incoming call blocking probability $P_{nc}$, is the fraction of the number of obstructed SU calls to the number of incoming SU calls. It can be expressed as
\begin{equation}
P_{nc}=\frac{\mbox{\textit{(Number of blocked SU calls/request)}}}{\mbox{\textit{(Number of new SU calls/request)}}}                    
\end{equation}
The offered traffic $\omega_ot$ is defined as the new call arrival rate divided by the average cell dwelling time. It is expressed as,  
\begin{equation}
\omega_ot=\frac{\lambda_{nc}}{\mu_d}
\end{equation}                                                         
Handover failure probability $P_{hf}$ is define as the fraction of SU calls been forced to terminate to the number of successful new SU calls. It can be expressed as,
\begin{equation}
P_{hf}=\frac{\mbox{(\textit{Number of  SU calls been fored to terminate)}}}{\mbox{(\textit{number of  successful new SU calls/request)}}}
\end{equation}

\section{Results and discussion}\label{sec7}
\begin{figure}[h]
	\centering
	\includegraphics[width=1\linewidth]{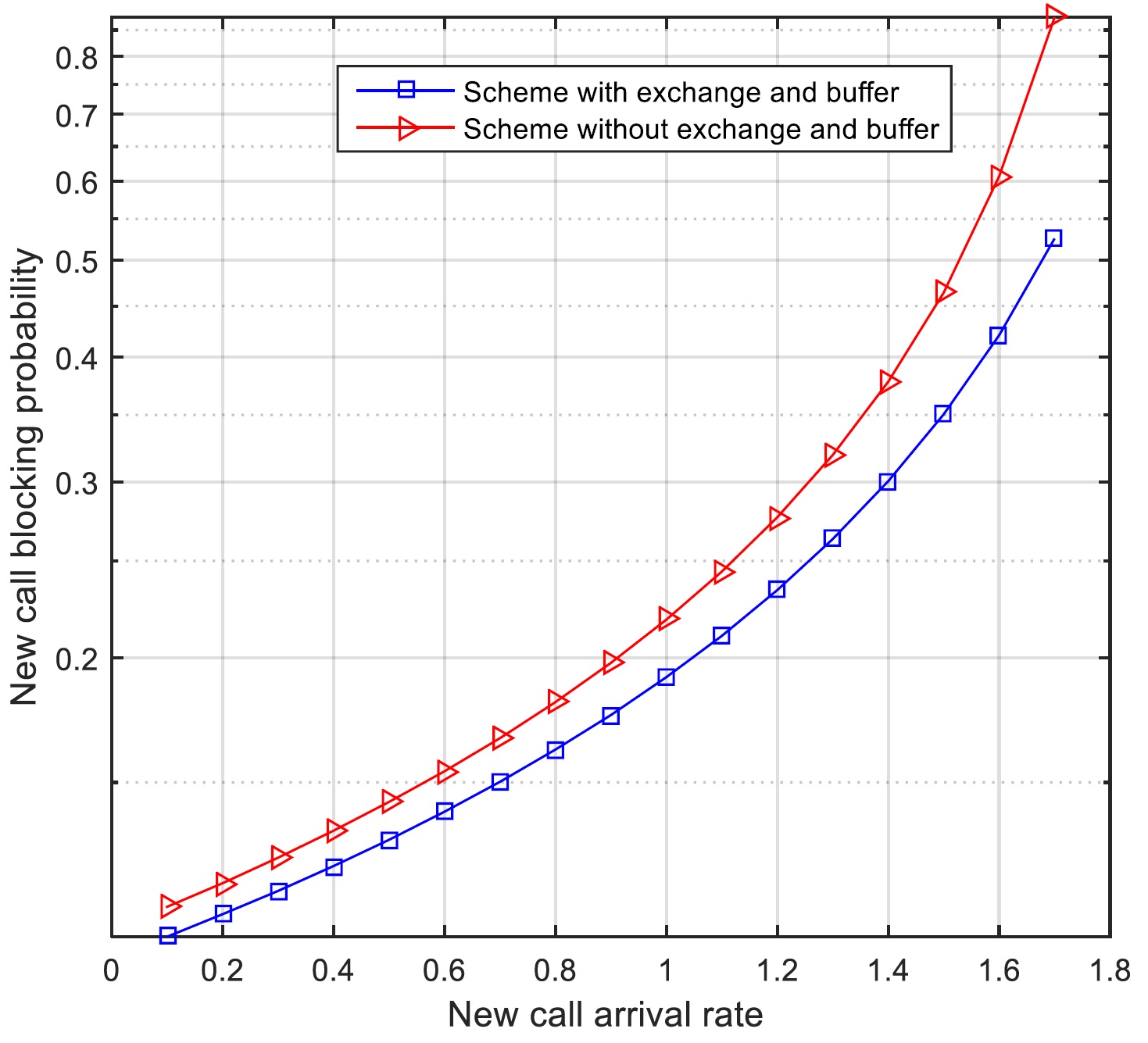}
	\caption{New call blocking probability vs. new call arrival rate.}
	\label{fig:fg4}
\end{figure}

The numerical results are presented from simulations run in Matlab platform. Table III shows the simulation parameters that were used. 
\begin{table}\label{tb3}
	\caption{Simulation parameters}
	\centering
\begin{tabular}{ll}
$\textit{Numbers of channels:}$&10\\
$\textit{Handover queue:}$&2\\
$\textit{New call arrival rate:}$&$1\textit{ call/sec}$\\
$\textit{Average calls residence time:}$& $120\textit{ sec}$\\
$\textit{New call arrival rate:}$&$1\textit{ call/sec}$\\
$\textit{Average calls residence time:}$&$120 \textit{ sec}$\\
$\textit{Average call holding time:}$&$240 \textit{ sec}$
\end{tabular}
\end{table}
In Fig. \ref{fig:fg4}, the two schemes are compared. From our result, as new calls arrive, the bocking probability increases as expected. However, the scheme with the handover exchange protocol has a better performance and with the buffer which accommodate request that has not been serviced.

\begin{figure}[h]
	\centering
	\includegraphics[width=1\linewidth]{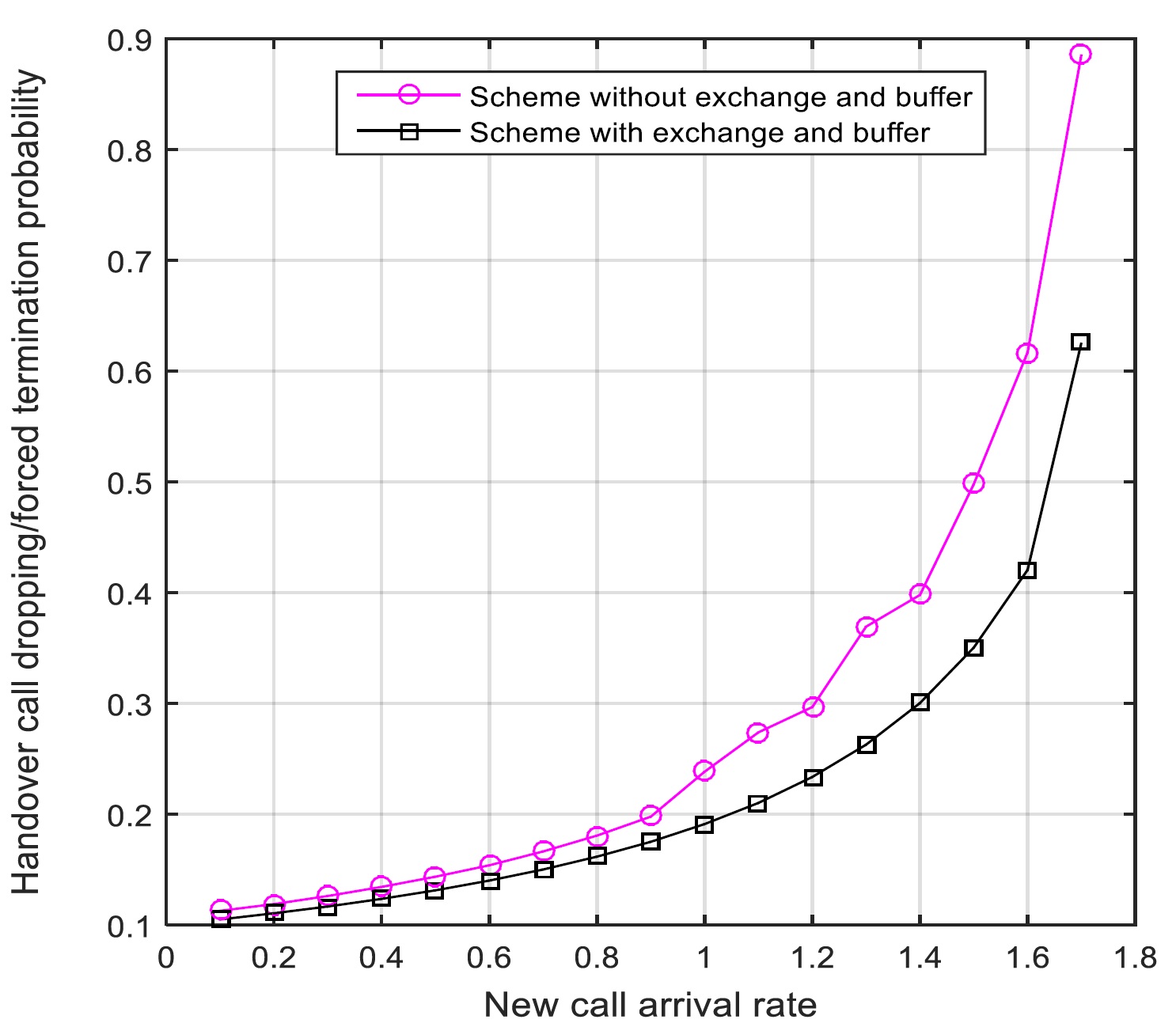}
	\caption{Dropping/forced termination probability vs. call arrival rate.}
	\label{fig:fg5}
\end{figure}
\begin{figure}[h]
	\centering
	\includegraphics[width=1\linewidth]{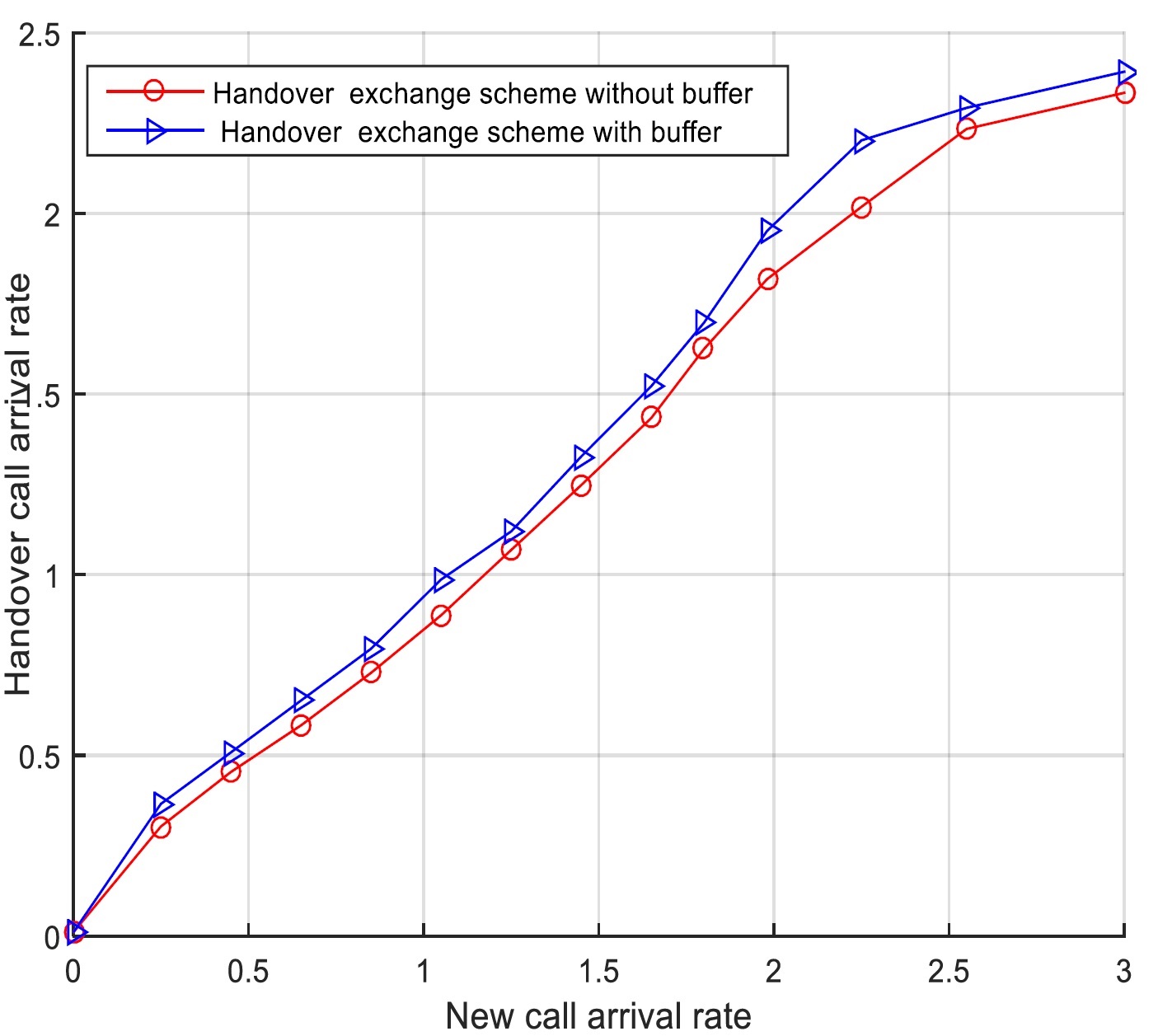}
	\caption{Handover call arrival rate vs. new call arrival rate.}
	\label{fig:fg6}
\end{figure}

Fig. \ref{fig:fg5} implies that once a request/call cannot be serviced, or queued it is dropped. On the other hand, when a call is being serviced and a more superior (high priority users) call comes in, if it cannot be queued, then it is forced to terminate. The difference is that for forced termination, access would have been granted. However, the scheme with both the buffer regime and exchange protocol outperformed the conventional scheme with these robust techniques. 
When a new call arrives from a user, due to mobility, there is tendency of handing over that call form one cell to another. However, if there is a buffer, the handover will be more successful in the sense that instead of dropping the call or request, it is queued in a buffer \cite{esenogho2017}, \cite{esenogho2015}. This has been illustrated in our result in Fig. \ref{fig:fg6}. 
 
\begin{figure}[h]
	\centering
	\includegraphics[width=1\linewidth]{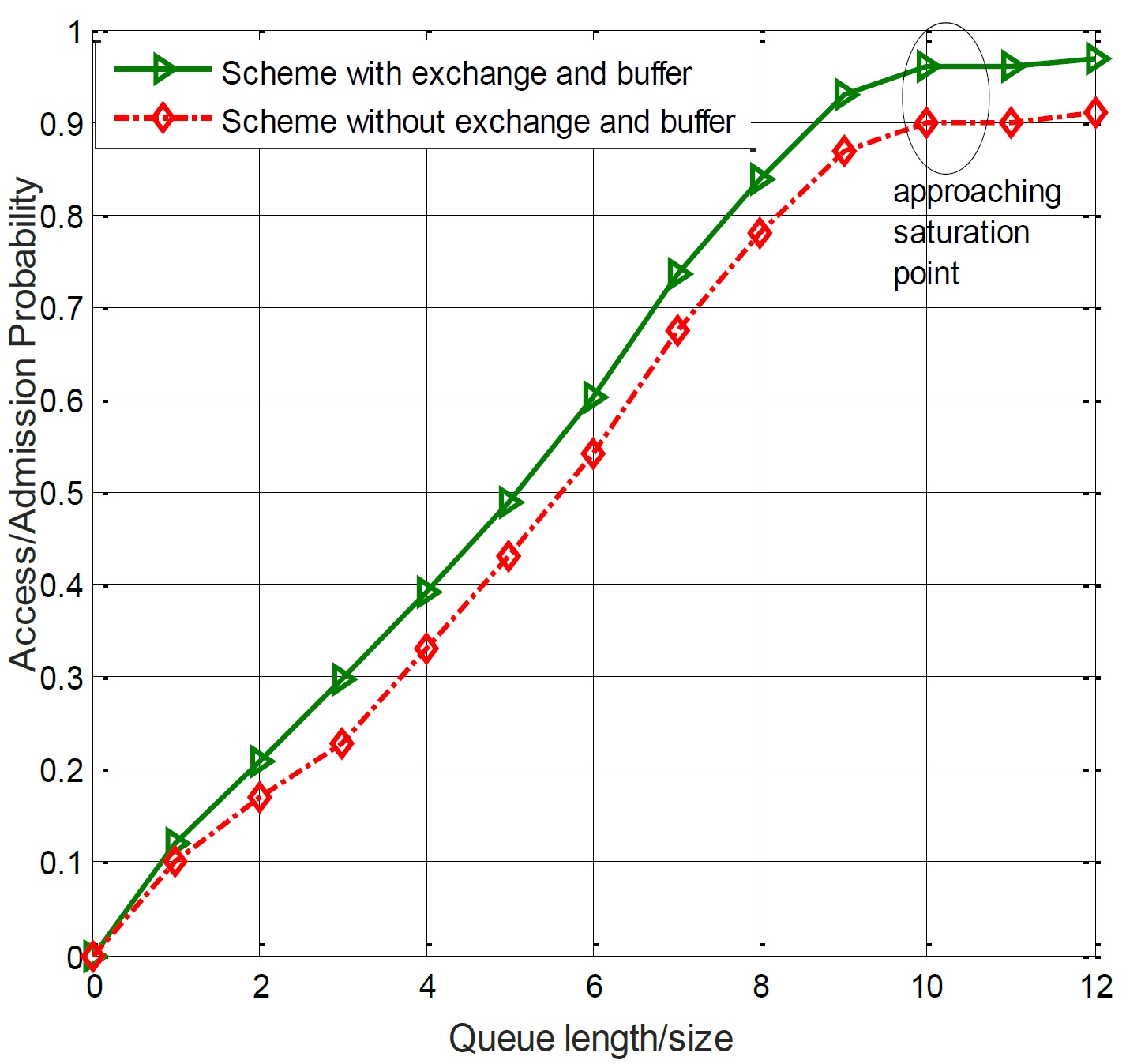}
	\caption{Access/Admission prob. vs. Queuing length.}
	\label{fig:fg7}
\end{figure}

Fig. \ref{fig:fg7} shows the impact of a queuing regime on the access or admission probability. This gives SUs the leverage (avenue to wait) to access the channel whenever it is interrupted or if the SU experiences insufficient or no resources channel. As the queue length increases, the probability of SU accessing the resources increases as well, and at a certain point, begins to saturate due fixed buffer capacity. However, due to space constrain, more results would have been presented but will be considered in future investigation.

\section{Conclusion}\label{sec:conclusion}
 This paper investigated through a comparative study the impact of a queuing regime of a handover exchange scheme. The role a buffer/queuing regime plays that makes handover a success (reduced handover failure) cannot be overemphasized especially when new calls arrive in batches. Our future work will include a detail performance analysis of the scheme investigated using Markovian model. Also, the class of secondary traffic/users will be considered.

\end{document}